\documentclass[12pt]{article}

\usepackage{setspace}
\usepackage{amsmath}
\usepackage{graphicx}
\usepackage{geometry}

\usepackage{array}
\usepackage{booktabs}
\usepackage[normalem]{ulem}

\ifx\pdfoutput\@undefined\usepackage[usenames,dvips]{color}
\else\usepackage[usenames,dvipsnames]{color}
\IfFileExists{pdfcolmk.sty}{\usepackage{pdfcolmk}}{}
\fi
\usepackage[plainpages=false,pdfpagelabels,pagebackref=false,naturalnames=true,hyperindex=true]{hyperref}
\hypersetup{colorlinks=true,
urlcolor=Cerulean,linkcolor=BrickRed,citecolor=RoyalBlue,
  pdfpagemode=UseNone,
  pdfstartview=FitH}
\usepackage[all]{hypcap}

\title{Emergence in artificial life}

\author{
Carlos Gershenson$^{1,2,3}$\\
\mbox{}\\
{\scriptsize $^1$Instituto de Investigaciones en Matem\'aticas Aplicadas y en Sistemas}\\
{\scriptsize \& Centro de Ciencias de la Complejidad, Universidad Nacional Aut\'{o}noma de M\'{e}xico, Mexico City, Mexico}\\
{\scriptsize $^2$Lakeside Labs GmbH, Klagenfurt, Austria}\\
{\scriptsize $^3$Santa Fe Institute. 1399 Hyde Park Rd., Santa Fe, NM 87501, USA}\\
{\scriptsize cgg@unam.mx}} 

\begin{document}
\maketitle

\begin{abstract}
Even when concepts similar to emergence have been used since antiquity, we lack an agreed definition. However, emergence has been identified as one of the main features of complex systems. Most would agree on the statement ``life is complex''. Thus, understanding emergence and complexity should benefit the study of living systems. 

It can be said that life emerges from the interactions of complex molecules. But how useful is this to understand living systems?
Artificial life (ALife) has been developed in recent decades to study life using a synthetic approach: build it to understand it. ALife systems are not so complex, be them soft (simulations), hard (robots), or wet (protocells). Then, we can aim at first understanding emergence in ALife, for then using this knowledge in biology.

I argue that to understand emergence and life, it becomes useful to use information as a framework. In a general sense, I define emergence as information that is not present at one scale but is present at another scale. This perspective avoids problems of studying emergence from a materialist framework, and can be also useful in the study of self-organization and complexity.
	
\end{abstract}



\section{Emergence}

The idea of emergence is far from new~\cite{Wimsatt1986}. It has certain analogies with Aristotle referring to the whole as being more than the sum of its parts. It had some development in the XIXth century, starting with John Stuart Mill, in what is known as ``British emergentism''~\cite{mclaughlin1992rise,Mengal2006}. However, given the success of reductionist approaches, interest on emergence diminished in the early XXth century.

Only in recent decades it has been possible to study emergence systematically, since we lacked the proper tools to explore models of emergent phenomena before digital computers were developed~\cite{Pagels1989}. 
 
In parallel, limits of reductionism have surfaced~\cite{Morin2006,HeylighenEtAl2007,Gershenson2013Facing-Complexi}. While successful in describing phenomena in isolation, reductionism has been inadequate for studying emergence and complexity. As Murray Gell-Mann put it: ``Reductionism is correct, but incomplete''.

This was already noted by Anderson~\cite{Anderson1972} and others, as phenomena at different scales exhibit properties and functionalities that cannot be reduced to lower scales~\cite{Gu2009}. Thus, the reductionist attempt of basing all phenomena in the ``lowest'' scale (or level) and declaring only that as reality, while everything else is epiphenomena, has failed miserably. Nevertheless, it still has several followers, as an coherent alternative has yet to emerge (pun intended). 

I blame the failure of reductionism on complexity~\cite{Bar-Yam1997,Mitchell:2009,ComplexityExplained,Ladyman2020}. Complexity is characterized by relevant \emph{interactions}~\cite{Gershenson:2011e}. These interactions generate novel \emph{information} that is not present in initial nor boundary conditions. Thus, predictability is inherently limited, due to \emph{computational irreducibility}~\cite{Wolfram:2002,Chaitin2013}: there are no shortcuts to the future, as information and computation produced during the dynamics of a system require to go through all intermediate states to reach a final state. The concept of computational irreducibility was already suggested by Leibniz in 1686~\cite{Chaitin2013},
but its implications have been explored only recently~\cite{Wolfram:2002}.
An integrated theory of complexity is lacking, but its advances have been enough to prompt the abandonment of the reductionist enterprise. I do not see the goal of complexity as fulfilling the expectations of a Laplacian worldview, where everything can be predictable only if we have enough information and computing power. On the contrary, complexity is shifting our worldview~\cite{Morin2006,HeylighenEtAl2007}, so that we are understanding the limits of science and seek not only prediction, but also adaptation~\cite{Gershenson2013Facing-Complexi}. Instead of attempting to dominate Nature for our purposes, we are learning to take our place in it.

Thus, we are slowly accepting emergence as \emph{real}, in the sense that emergent properties have causal influence in the physical world (see below concerning downward causation). Nevertheless, we still lack an agreed definition of emergence~\cite{FeltzEtAl2006,bedau2008emergence}. This might seem problematic, but we also lack agreed definitions of complexity, life, intelligence, and consciousness. However, this has not prevented advances in complex systems, biology, and cognitive sciences. 

In a general sense, we can understand emergence as \emph{information that is not present at one scale but is present at another scale}. For example, life is not present at the molecular scale, but it is at the cell and organism scales. When scales are spatial or organizational, emergence can be said to be \emph{synchronic}, while emergence can be \emph{diachronic} for temporal scales~\cite{Rueger2000}. 

It could be argued that the above definition of emergence is not sharp. I do not believe that emergence --- similar to life --- \emph{is} or \emph{is not}. We should speak about different degrees of emergence, and this can be useful to compare ``more'' or ``less'' emergence in different conditions and systems. Another argument against the definition could be that of vagueness. Still, sharp definitions tend to be useful only for particular contexts. This definition is general enough to be applicable in a broad variety of contexts, and particular, sharper notions of emergence can be derived from it for specific situations, \emph{e.g.}, ~\cite{cooper2009emergence,Neuman2019,doi:10.1098/rsta.2020.0429}.

There have been different types of emergence proposed, but we can distinguish mainly \emph{weak} emergence and \emph{strong} emergence. Weak emergence~\cite{Bedau:1997} only requires computational irreducibility, so it is easier to accept for most people. The ``problem'' with strong emergence~\cite{BarYam:2004} is that it implies \emph{downward causation}~\cite{Campbell1974,Bitbol2012,Flack2017,farnsworth_ellis_jaeger_2017}.

Usually, emergent properties are considered to occur at higher/slower scales, arising from the interactions occurring at lower/faster scales. Still, I argue that emergence can also arise in lower/faster scales from interactions at higher/slower scales, as exemplified by downward causation. This is also related to ``causal emergence''~\cite{Hoel19790,e19050188}. Taking again the example of life, the organization of a cell restricts the possibilities of its molecules~\cite{Kauffman2000}. Most biological molecules would not exist without life to produce them. Molecules in cells do not violate the laws of physics nor chemistry, but these are not enough to describe the existence of all molecules, as information (and emergence) may flow across scales in either direction.

Note that not all properties are emergent. Only those that require more than one scale to be described (thus the novel information). For example, in a crowd, there might be some emergent properties (\emph{e.g.}, coherent ``Mexican wave'' in a stadium.), but not all of the crowd properties are usefully described as emergent (\emph{e.g.}, traffic flow at low densities, as it can be described fully from the behavior of drivers). In the same sense, a society can produce emergent properties in its individuals (\emph{e.g.}, social norms that guide or restrict individual behavior), but not all properties of individuals are necessarily described as emergent in this (downward) way (\emph{e.g.} performance during a workout).
Complexity occurs when novel information is produced through \emph{interactions} between the components of a system. Emergence occurs when novel information is produced across \emph{scales}.

One might wonder whether then all macroscopic properties are emergent. No. If they can be fully derived from a microscopic description (information), then we can call them differently for convenience, but they can be practically reduced to the information of the microscale. If novel information is produced at the macroscale, then that information can be said to be emergent. 

Note that it is difficult to decide on the ontological status of emergence, \emph{i.e.}, whether it ``really exists'' (independently of an observer). I am aiming ``just'' for an epistemology of emergence, which can be understood as answering ``\emph{when is it useful to describe something as emergent?}''. The answer to this question can certainly change with contexts, so something might be usefully considered emergent in one context, but not in another one.

Since emergence is closely related to information, I will explain more about their relationship in Section~\ref{sec:Info}, but before, I will discuss the role of emergence in artificial life.

\section{(Artificial) Life}

There are several definitions of life, but none that everyone agrees with~\cite{Bedau2008What-is-Life,Zimmer2021}. We could simply say ``life is emergent'', but this does not explain much. 
Still, we can abstract the substrate of living systems and focus on the organization and properties of life. This was done already in cybernetics~\cite{HeylighenJoslyn2001,Gershenson2013The-Past-Presen}, but became central in artificial life~\cite{langton1997artificial,Adami:1998}. Beginning in the mid-1980s, ALife has studied living systems using a synthetic approach: building life to understand it better~\cite{Aguilar2014The-Past-Presen}. By having more precise control of ALife systems than in biological systems, we can study  emergence in ALife, increasing our general understanding of emergence. With this knowledge, we can go back to biology. Then, emergence might actually become useful to understand life.

ALife and its methods have had a considerable influence in cognitive sciences~\cite{Froese2010,doi:10.1177/1059712319856882,BeerInPress}. It still has to have an explicit impact in biology, perhaps because biological life is more complex than ALife, and biologists tend to study living systems directly. Still, computational models in biology are more and more commonplace \cite{noble2002rise}, so it could be said that the methods developed in artificial life have been absorbed in biology \cite{doi:10.1063/1.5038337} and other disciplines~\cite{lazer2009life,Rahwan2019,Trantopoulos2011,doi:10.1177/1356389020976153,Seth2021}.

Since one of the central goals of ALife is to understand the properties of living systems, it does not matter whether these are software simulations, robots, or protocells (representatives of ``soft'', ``hard'', and ``wet'' ALife)~\cite{Gershenson2020}. These approaches allow us to explore the principles of a ``general biology''~\cite{Kauffman2000} that is not restricted to the only life we know, based on carbon, DNA, and cells. 

\subsection{Soft ALife}

Mathematical and computational models of living systems have the advantage and disadvantage of simplicity: one can abstract physical and chemical details and focus on general features of life. 

A classical example is Conway's Game of Life~\cite{BerlekampEtAl82}. In this cellular automaton, cells on a grid interact with their neighbors to decide on the life or death of each cell. Even when rules are very simple, different patterns emerge, including some that exhibit locomotion, predation, and one could even say cognition~\cite{Beer2014}. Cells interact to produce higher-order emergent structures, that can be used to build logic gates, and even universal computation.

Another popular example is that of ``boids''~\cite{reynolds87flocks}: particles follow simple rules depending on their neighbors (try not to crash, try to keep average velocity, try to keep close). The interactions lead to the emergence of patterns similar to those of flocking, schooling, herding, and swarming. There have been several other models of collective dynamics of self-propelled agents~\cite{Sayama2008Swarm-Chemistry,Vicsek2012}, but the general idea is the same: local interactions lead to the emergence of global patterns.

Soft ALife has also been used to study open-ended evolution (OEE)~\cite{standish2003open,taylor2016open,Adams2017,pattee2019evolved}. For example, 
~\cite{hernandez2018undecidability} showed that undecidability and irreducibility are conditions for OEE. I would argue that OEE is an example of emergence, but not vice versa. Under the broader notion of emergence used in this paper, undecidability and irreducibility are not conditions for emergence.

Emergence in soft ALife perhaps is the easiest to observe, precisely because of its abstract nature. Even when most examples deal with ``upwards emergence'', there are also cases of ``downwards emergence'', \emph{e.g.},~\cite{Hoel19790,Escobar2019}, where information at a higher scale leads to novel properties at a lower scale. 

\subsection{Hard ALife}

One of the advantages and disadvantages of robots is that they are embedded and situated in a physical environment. The positive side is that they are realistic, and thus can be considered closer to biology than soft ALife. The negative side is that they are more difficult to build and explore. 

Emergence can be observed at the individual robot level, where different components interact to produce behavior that is not present in the parts, \emph{e.g.},~\cite{walter1950,walter1951machine,Braitenberg:1986}, or also at the collective level, where several robots interact to achieve goals that individuals are unable to fulfill~\cite{DorigoEtAl2004,ZykovEtAl2005,Halloy2007,RubensteinEtAl2014,Vasarhelyi2018}.

Understanding emergent properties of robots and their collectives is giving us insights into the emergent properties of organisms and societies. And as we better understand organisms and societies, we will be able to build robots and other artificial systems that exhibit more properties of living systems~\cite{Bedau:2009,Bedau2013IntroductionLT,Gershenson:2013}.

\subsection{Wet ALife}

The advantage and disadvantage of wet ALife is that it deals directly with chemical systems to explore the properties of living systems. By using chemistry, we are closer to biological life than with soft or hard ALife. However, the potential space of chemical reactions is so huge, and its exploration is so slow, that it seems amazing that there have been any advances at all using this approach. 

One research avenue in wet ALife is to attempt to build ``protocells''~\cite{rasmussen2003bridging,hanczyc2003experimental,rasmussen2004transitions,Protocells2008}: chemical systems with some of the features of living cells, such as membranes, metabolism, information codification, locomotion, reproduction, etc. Dynamic formation and maintenance of micelles and vesicles \cite{luisi1989self,bachmann1990self,bachmann1992autocatalytic,walde1994autopoietic} predate the protocell approach, while more recently, the properties of active droplets or ``liquid robots'' \cite{vcejkova2017droplets} have been an intense area of study. These include the emergence of locomotion~\cite{hanczyc2007fatty,cejkova2014dynamics} and complex morphology~\cite{vcejkova2018multi}.

There have also been recent works studying collective properties of protocells~\cite{Qiao:2017} and droplets~\cite{vcejkova2017droplets}, where the interactions between the chemical entities lead to the emergence of global patterns.

The recent development of ``xenobots'' \cite{Kriegman2020,Blackistoneabf1571,blackiston2022biological} --- multicellular entities designed using artificial evolution and constructed from frog embryonic cells --- can also be considered as wet ALife.

\section{Information}
\label{sec:Info}

Our species has lived through three major revolutions: agricultural, industrial, and informational. We can say that the first one dealt mainly with the control of matter, the second with control of energy, and the third and current one with the control of... information, obviously. This does not mean that we did not manipulate information beforehand~\cite{gleick2011information}, but that we lacked the tools to do so at the scales we have since the development of electronic computers.

Shannon~\cite{Shannon1948} proposed a measure of information in the context of telecommunications. This has been useful, but also many sophistications have been derived from it. Shannon's information can be seen as a measure of a ``just so'' arrangement, so it can also be used to measure organization. Still, it is ``simply'' a probabilistic measure that assumes the meaning of a message is shared by sender and receiver. But of course, the same message can acquire different meanings depending on the encoding used~\cite{Haken2015Information-Ada}.

Living systems process information~\cite{Hopfield1994,Farnsworth2013Living-is-Infor}. Thus, understanding information might improve our understanding of life~\cite{Kim2021}. It has been challenging to describe information in terms of physics (matter and energy)~\cite{Kauffman2000}, especially when we are interested in the \emph{meaning} of information~\cite{Neuman:2008,Haken2015Information-Ada,Scharf2021}. 

One alternative is to describe the world in terms of information, including matter and energy~\cite{Gershenson:2007}. Everything we perceive can be described in terms of information: particles, fields, atoms, molecules, cells, organisms, virus, societies, ecosystems, biospheres, galaxies, \emph{etc}, simply because we can name them. If we can name them, then they can be described as information. If we could not name them, then we should not even speak about them~\cite{wittgenstein2013tractatus}. All of these have physical components. Nevertheless, other non-physical phenomena can be also described in terms of information, such as interactions, ideas, values, concepts, and money. This gives information the potential to bridge between physical and non-physical phenomena, avoiding dualisms. This does not mean that other descriptions of the world are ``wrong''. One can have different, complementary descriptions of the same phenomenon, and this does not affect the phenomenon. The question is how useful are these descriptions for a particular purpose. I claim that information is useful to describe general principles of our universe, as an information-based formalism can be applied easily across scales. Thus, general ``laws'' of information can be explored, generalizing principles from physics, biology, cybernetics, complexity, psychology, and philosophy~\cite{Gershenson:2007}. These laws can be used to describe and generalize known phenomena within a common framework.
Moreover, as von Baeyer~\cite{vonBaeyer2004} suggested, information can be used as a language to bridge across disciplines.

One important aspect of information is that it is not necessarily conserved (as matter and energy). Information can be created, destroyed, or transformed. We can call this \emph{computation}~\cite{Denning:2010,Gershenson:2010b}. I argue that some of the ``problems'' of emergence arise due to conservation assumptions, that dissolve when we describe phenomena in terms of information. For example, meaning can change passively~\cite{Gershenson:2007}, \emph{i.e.}, independently of its substrate. One instance of this is the devaluation of money: prices might change, while the molecules of a bill or the atoms of a coin remain unaffected by this. 

There have been several measures of emergence proposed, \emph{e.g.},~\cite{Bersini2006,Prokopenko:2008,Fuentes2014}. In the context posed in this paper, it becomes useful to explore a notion of emergence in terms of information, since it can be applied to everything we perceive.

We have proposed a measure of emergence that is actually equivalent to Shannon's information~\cite{Shannon1948,Fernandez2013Information-Mea} (which is also equivalent to Boltzmann-Gibbs entropy). Shannon was interested in a function to measure how much information a process ``produces". Thus, if we understand emergence as ``new'' information, we can measure emergence $E$ (diachronic or synchronic, weak or strong) as well with information:

\begin{equation}
E = -K \sum_{i=i}^{n} p_{i} \log p_{i}, 
\end{equation}

where $K$ is a positive constant that can be adjusted to normalize $E$ to the interval $[0,1]$ depending on the ``alphabet'' of length $n$. If we use $\log_{2}$, then 

\begin{equation}
K = \frac{1}{\log_{2}n}.
\end{equation}

This normalization allows us to use the same measure, and explore different information is produced at different scales. If $E=0$, then there is no new information produced: we know the future from the past, one scale from another. If $E=1$, we have maximum information produced: we have no way of knowing the future from the past, or one scale from another; we have to observe them.

In this context, ``new information'' does not imply something that has never been produced before, but a pattern that deviates from the probability distribution of previous patterns: if previous information tells you nothing about future information (as in the case of fair coin tosses), then each symbol will bring maximum information. If future symbols can be predicted from the past (which occurs when only one symbol has maximum probability of occurring and all the others never occur), then these ``new'' symbols carry no information at all.

If we already have information at one scale, but observe ``new'' information at another scale, \emph{i.e.}, that cannot be derived from the information at the first scale, then we can call this information emergent. 

This measure of emergence has been successfully been applied in different contexts~\cite{Zubillaga2014Measuring-the-C,Febres2013Complexity-meas,Amoretti2015Measuring-the-c,FernandezCxLakes,10.3389/fphy.2018.00045,10.3389/fphy.2020.00035,e22010089,Ramirez-Carrillo2020,Zapata2020}. Still, this does not imply that other measures of emergence are ``wrong'' or not useful, as usefulness depends on the contexts in which a measure is applied~\cite{Gershenson2002ua}. Moreover, the goal of this paper lies not on defending a measure of emergence, but on exploring the concept of emergence.

A better understanding of emergence has been and can be useful for soft, hard, and wet ALifes. In all of them, we are interested on how properties of the living emerge from simpler components. In less cases, we are also interested on how systems constrain and promote behaviors and properties of their components (downward emergence). 

Emergence is problematic only in a physicalist, reductionist worldview. In an informational, complex worldview, emergence is natural to accept. Interactions are not necessarily described by physics, but they are “real”, in the sense that they have causal influence on the world. We can use again the example of money. The value of money is not physical, it is informational. The physical properties of shells, seeds, coins, bills, or bits do not determine their value. This is very clear with art. The value comes from the interactions among humans who agree on it. The transformation of a mountain excavated for open-pit mining does not violate the laws of physics. But, using only the laws of physics, one cannot predict that humans might decide to give value to some mineral under the mountain and transform matter and energy to extract it. In this sense, information (interactions, money) has a (downward) causal effect on matter and energy.


Using an informationist perspective, one also avoids problems with downward causation, as this can be seen as simply the effect of a change of scales~\cite{BarYam2004}. In the Game of Life, one can describe gliders (higher scale) as emerging from cell rules (lower scale), but also describe cell states as emerging from the movement of the glider on its environment. In a biological cell (higher scale), one can describe life as emerging from the interactions of molecules (lower scale), but also as molecule states and behavior emerging from the cell's constraints. In many cases, biological molecules would simply degrade if they were not inside an organism that produces and provides the conditions for sustaining them. In a society (higher scale), one can describe culture and values as emerging from the interactions of people (lower scale), but also to describe individual properties and behaviors as emerging from social norms. 

Like with special and general relativity in physics, one cannot define one ``real'' scale of observation (frame of reference). Scales are relative to an observer, just as the information perceived at each scale. \emph{Information is relative to the agent perceiving it}. In other words, as mentioned above, different meanings can be implied by the same messages.

\section{Self-organization}

Emergence is one of the main features of complex systems~\cite{ComplexityExplained}. Another is self-organization~\cite{Ashby1947sos,Ashby1962,AtlanCohen1998,Heylighen2003sos,GershensonHeylighen2003a}, and it has also had a great influence in ALife~\cite{Gershenson2020}. A system can be usefully described as self-organizing when global patterns or behaviors are a product of the interactions of its components~\cite{GershensonDCSOS}.

One might think that self-organization requires emergence, and vice versa; or at least, that they go hand in hand. However, self-organization and emergence are better understood as opposites~\cite{LopezRuiz:1995,Fernandez2013Information-Mea}. 

Self-organization can be seen as an increase in order. This implies a reduction of entropy~\cite{GershensonHeylighen2003a}. If emergence leads to an increase of information, which is analogous to entropy and disorder, self-organization should be anti-correlated with emergence, as more organization requires less information. We can thus simply measure self-organization $S$ as:

\begin{equation}
S = 1 - E.
\label{eq:S}
\end{equation}

This equation assumes that the dynamics are internal, so that the organization is self-produced. Otherwise, we might be measuring ``exo-organization''.  Minimum $S=0$ occurs for maximal entropy: there is only change. In a system where all of its states have the same probability, there is no organization. Maximum $S=1$ occurs when order is maximum: there is no change. Only one state is possible, and we can call this state ``organized''~\cite{Ashby1962}. Note that this measure is not useful to decide whether a system is self-organizing or not~\cite{GershensonHeylighen2003a}. The purpose of the measure is to compare different levels of organization in a specific context.

Also, it should be noted that information (and thus emergence and self-organization as considered here) can have different dynamics at different scales. For example, there can be more self-organization at one scale (spatial or temporal) and more emergence at another scale.

\section{Complexity}

\begin{figure}[htbp]
\begin{center}
\includegraphics[scale=0.65]{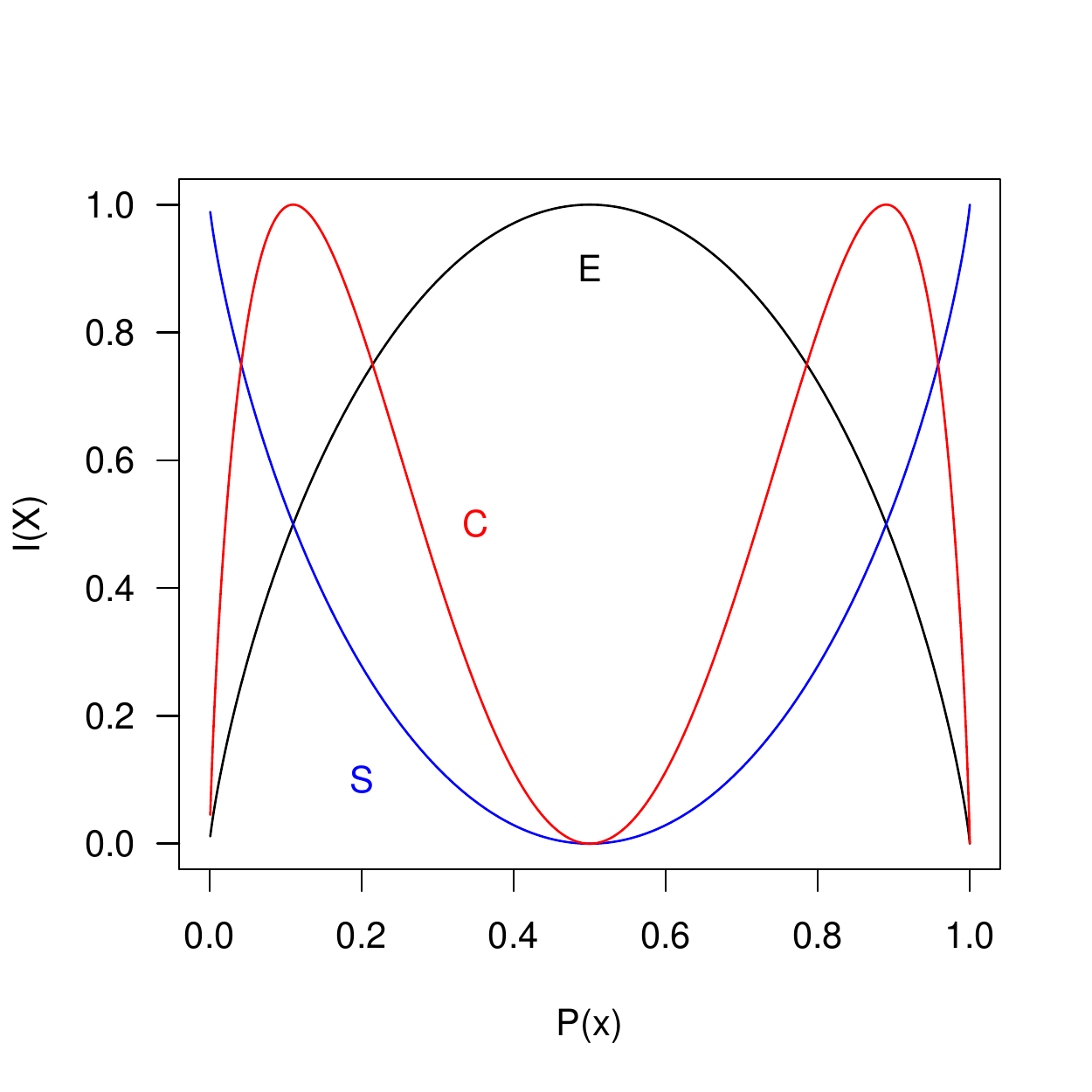}
\caption{Emergence $E$, self-organization $S$, and complexity $C$ depending on the probability of having ones $P(x)$ in a binary string~\cite{Fernandez2013Information-Mea}.}
\label{fig:ESC}
\end{center}
\end{figure}

As mentioned above, complex systems tend to exhibit both emergence and self-organization. Extreme emergence implies chaos, while extreme self-organization implies immutability (order). Complexity requires a \emph{balance} between both emergence and self-organization~\cite{LopezRuiz:1995}. Therefore, we can measure complexity $C$ with:

\begin{equation} 
C = 4 \cdot E \cdot S,
\label{eq:C}
\end{equation}

where the 4 is added to normalize $C$ to $[0,1]$. $C$ will be close to zero if either emergence or self-organization dominate, and it will increase as these become more balanced (see Figure~\ref{fig:ESC}). 

This measure $C$ of complexity is maximal at phase transitions in random Boolean networks~\cite{GershensonFernandez:2012}, the Ising model, and other dynamical systems characterized by criticality~\cite{Febres2013Complexity-meas,Zubillaga2014Measuring-the-C,Amoretti2015Measuring-the-c,Ramirez-Carrillo2018,Pineda2019,Franco2021}. Recently, we have found that different types of heterogeneity increase the parameter regions of high complexity for a variety of models~\cite{Sanchez-Puig2022,lopez2022temporal}. 

Interestingly, typical examples of emergence and self-organization are not extreme cases of either. Perhaps this is the case because if we had only emergence or only self-organization, then these would not be distinguishable from the full system. It is easier provide examples in contrast to another property. So, \emph{e.g.}, a flock of birds is a good example of emergence, self-organization and complexity, because the flock produces novel information, self-organizes, \emph{and} has interactions at the same time. If it had $E=1, S=0, C=0$, then only information would be produced constantly (no complexity nor organization). If the flock had $E=0, S=1, C=0$, then it would be static, fully organized, without change (no complexity nor emergence).

\section{Discussion}

Emergence is partially subjective, in the sense that the ``emergentness'' of a phenomenon can change depending on the frame of reference of the observer. This is also the case with self-organization~\cite{GershensonHeylighen2003a} and complexity~\cite{BarYam2004}. Actually, anything we perceive, all information, might change with the \emph{context}~\cite{Gershenson2002ua} in which it is used. Of course, this does not mean that one cannot be objective about emergence, self-organization, complexity, life, cognition~\cite{Gershenson2004}, and so on. We just have to agree on the context (frame of reference) first.

Therefore, the question is not whether something \emph{is} emergent or not. The question becomes: in which contexts it is \emph{useful} to describe something as emergent? If the context just focusses on one scale, it does not make sense to speak of emergence. But if the context implies more than one scale, and how phenomena/information at one scale affects phenomena/information at another scale, then emergence becomes relevant.

Thus, is emergence an essential aspect of (artificial) life? It depends. If we are interested in life at a single scale, we can do without emergence. But if we are interested on the relationships across scales in living systems, then emergence becomes a \emph{necessary} condition for life: life \emph{has} to be emergent if we are interested in explaining living systems from non-living components. Without emergence, we would fall into dualisms. A similar argument can be made for the study of cognition.

Moreover, information has been proposed to measure how ``living'' a system might be~\cite{Fernandez2013Information-Mea,Farnsworth2013Living-is-Infor,Kim2021}. This view also suggests that there is no sharp transition between the non-living and living, but rather a gradual increase on how much information is produced by an organism compared to how much of its information is produced by its environment~\cite{Gershenson:2007}.

One implication is that materialism becomes insufficient to study life, artificial or biological. Better said: materialism was always insufficient to study life. Only now we are developing an alternative. It remains to be seen whether it is a better one.

There are inherent limitations to formal systems~\cite{godel1931formal,Turing:1936}. These limitations also apply to artificial intelligence~\cite{mitchell2019artificial} and soft and hard ALife. Simply described: a system cannot change its own axioms. One can always define a metasystem, where change will be possible (in a predefined way), but there will be new axioms which will not be possible to change.
Since scientific theories are also formal, they are also limited in this way. This is one of the reasons for why emergence and downward causation are difficult to accept for some. It implies that there is no hope of a grand unified theory, as the emergent future cannot be prestated~\cite{Kauffman2008} and downward causation can change axioms of our theories. Thus, the traditional approach has been to deny emergence and downward causation. I believe that this is untenable~\cite{Wolpert2022}, and that we have to develop a scientific understanding of phenomena that --- even when we know it cannot be complete --- can be always evolving.

\section*{Acknowledgements}

I am grateful to Edoardo Arroyo, Manlio De Domenico, Nelson Fern\'andez, Bernardo Fuentes Herrera, Hiroki Sayama, Vito Trianni, Justin Werfel, and David Wolpert for useful discussions. Anonymous referees provided comments that improved the paper. I acknowledge support from UNAM-PAPIIT (IN107919, IV100120, IN105122) and from the PASPA program from UNAM-DGAPA.

\bibliographystyle{alj}
\bibliography{refs,carlos}

\end{document}